\documentclass[superscriptaddress,aps,journal=prl,preprint]{revtex4-2}
\usepackage{graphicx}
\usepackage{epstopdf}
\usepackage{makecell}
\usepackage{booktabs}
\usepackage{array}
\usepackage{url}
\usepackage{bm,amsmath,amssymb} 
\usepackage{color, soul} 
\usepackage{dcolumn}   
\usepackage{multirow}

\newcommand{\etal}{{\em et al.\ }}
\newcommand{\comment}[1]{}

\newcommand{\ie}{\textit{i}.\textit{e}.}

\begin{document}

\title{Origin of ferroelectricity in magnesium doped zinc oxide}
\author{Jiawei Huang}
\thanks{These two authors contributed equally}
\affiliation{Zhejiang University, Hangzhou, Zhejiang 310058, China}
\affiliation{Key Laboratory for Quantum Materials of Zhejiang Province, Department of Physics, School of Science, Westlake University, Hangzhou Zhejiang 310024, China}
\author{Yihao Hu}
\thanks{These two authors contributed equally}
\affiliation{Zhejiang University, Hangzhou, Zhejiang 310058, China}
\affiliation{Key Laboratory for Quantum Materials of Zhejiang Province, Department of Physics, School of Science, Westlake University, Hangzhou Zhejiang 310024, China}
\author{Shi Liu}
\email{liushi@westlake.edu.cn}
\affiliation{Key Laboratory for Quantum Materials of Zhejiang Province, Department of Physics, School of Science, Westlake University, Hangzhou Zhejiang 310024, China}
\affiliation{Institute of Natural Sciences, Westlake Institute for Advanced Study, Hangzhou, Zhejiang 310024, China}

\begin{abstract}{ 
Recent experiments demonstrated robust ferroelectricity in Mg-doped ZnO (ZMO) of the wurtzite structure, hinting at a promising strategy to substantially expand the list of ferroelectrics by doping conventional piezoelectrics. We investigate the origin of ferroelectricity in ZMO with first-principles density functional theory (DFT). The general argument that the Mg alloying could soften the ionic potential energy surface of ZMO for polarization reversal is overly simplified. Our DFT calculations reveal that even at a high Mg concentration, the energy difference ($\Delta U$) between the polar and nonpolar phases remains prohibitively large for ZMO systems when the strain is fixed to the polar phase. Interestingly, the magnitude of $\Delta U$ becomes substantially smaller when the strain relaxation is allowed, approaching the value of typical perovskite ferroelectrics such as PbTiO$_3$ with increasing Mg doping concentrations. The enabled switchability of ZMO systems is attributed to a hexagonal phase of MgO that is much lower in energy than its wurtzite counterpart. Detailed orbital and bonding analysis supports that the intra-atomic $3d_{z^2}$-$4p_z$ orbital self-mixing of Zn plays an important role in stabilizing the polar wurtzite phase, the lack of which is responsible for the low-energy nonpolar hexagonal phase of MgO.
}
\end{abstract}

\maketitle
\newpage




\section{Introduction}
Ferroelectrics afford two or more orientational states, each associated with a spontaneous electric polarization, and can be switched between these states by applying an electric field. The polarization is temperature dependent, and the switching of the polarization is often coupled with the strain change of the material. Therefore, all ferroelectrics are pyroelectrics/piezoelectrics but not vice versa as the polarization of some pyroelectrics/piezoelectrics are not reversible~\cite{Lines77}. Ferroelectric perovskite oxides of $AB$O$_3$ chemical formula are perhaps the most studied and used ferroelectric materials~\cite{Rabe07Book}. Because $A$- and $B$-sites can accommodate a wide range of elements, the structural and electronic properties of ferroelectric perovskites are highly tunable, capable of supporting a wealth of properties going beyond ferroelectricity such as ferroelasticity~\cite{Lee06p214110}, ferromagnetism~\cite{Sakai11p137601,Hayashi11p12547}, multiferroicity~\cite{Hill00p6694, Wang03p1719}, and superconductivity~\cite{Baumert95p175,Rischau17p643}. However, the structural and chemical complexity of perovskite ferroelectrics makes it challenging to integrate these materials into a semiconductor manufacturing process that often has stringent requirements on the thermal budget and elements allowed in the production line~\cite{Mikolajick21p100901}. The poor compatibility of typical perovskite ferroelectrics such as Pb(Zr, Ti)O$_3$ with the complementary metal-oxide-semiconductor (CMOS) technology has been a major hurdle to downscale ferroelectric-based electronics such as ferroelectric memory to the sub-100~nm domain~\cite{TrolierMcKinstry04p7,Mikolajick20p1434}. 

The discovery of ferroelectricity in doped HfO$_2$~\cite{Boscke11p102903} and Al$_{1-x}$Sc$_x$N~\cite{Fichtner19p114103} thin films opened up exciting opportunities to incorporate a plethora of ferroelectric functionalities into integrated circuits because the parent materials, HfO$_2$ and AlN, have industry-validated compatibility with generic semiconducting technology~\cite{Gutowski02pB3.2, Stefon09p402}. The structural origin of ferroelectric HfO$_2$ is generally attributed to the metastable phase of space group $Pca2_1$~\cite{Materlik15p134109,Sang15p162905,Sebastian14p140103,Huan14p064111,Park15p1811} in thin films stabilized by various extrinsic factors including doping~\cite{Schroeder14p08LE02,Starschich17p333,Park17p4677,Xu17p124104,Batra17p9102}, surface/interface energy~\cite{Materlik15p134109,Polakowski15p232905,Park17p9973,Park15p1811}, clamping strain from capping electrodes~\cite{Batra17p4139,Shiraishi16p262904}, and oxygen vacancy~\cite{Xu16p091501,Pal17p022903}, though recent experimental and theoretical studies highlighted the importance of kinetic effects of phase transitions during the growth process~\cite{Park18p716,Park18p1800522,Liu19p054404}. By substituting Sc into AlN, a switchable polarization has been demonstrated in Al$_{1-x}$Sc$_x$N for $x>0.27$~\%, with both remnant polarization (80--110 $\mu$C/cm$^2$) and coercive fields (1.8--5~MV/cm) depending on $x$~\cite{Fichtner19p114103}. These findings, particularly the demonstrated polarization reversal in Al$_{1-x}$Sc$_x$N thin films, offer a fresh perspective for the search of ferroelectrics 
that are compatible with semiconductor process integration: ``soften" silicon-compatible piezoelectrics to make them switchable by applying appropriate chemical and/or physical ``stressors", hereinafter referred to as ``ferroelectrics everywhere" hypothesis~\cite{Ferri21p044101}.

Zinc oxide in the wurtzite structure ($wz$-ZnO) is a technologically important polar semiconductor with a direct band gap of 3.7~eV~\cite{Vogel95p14316} and a large electromechanical piezoelectric coefficient ($d_{33}=12.4$~pC/N)~\cite{Jang12p73503}. Various nanostructures of ZnO such as nanobelts, nanorings, nanowires, and nanocages~\cite{Wang04p26} have been synthesized, creating a rich platform for the design and development of nanoscale electronics. It was long postulated that $wz$-ZnO could be made ferroelectric through doping~\cite{Ozgur05p41301}. 
Onodera~\etal reported ferroelectricity in Zn$_{1-x}B_x$O ($B$ = Li, Mg), but the polarization values were unusually small (0.05--0.59 $\mu$C/cm$^{-2}$)~\cite{Onodera97p6008}, nearly three orders of magnitude smaller than ZnO ($\sim$90 $\mu$C/cm$^{-2}$). More recently, Ferri~\etal reported a giant switchable polarization of $>100$~$\mu$C/cm$^{-2}$ in Zn$_{1-x}$Mg$_{x}$O (ZMO) with $x$ in the range between $\sim$30\% and $\sim$37\% and coercive fields below 3~MV/cm at room temperatures, serving as a convincing example of ``ferroelectrics everywhere"~\cite{Ferri21p044101}. In comparison, conventional ferroelectrics like Pb(Zr,Ti)O$_3$ have a remnant polarization of 10--40~$\mu$C/cm$^{2}$ and a coercive field of 50--70 kV/cm~\cite{Budd85p107}; SrBi$_2$Ta$_2$O$_9$, being fatigue-free on Pt electrodes, has a smaller polarization of 5--10~$\mu$C/cm$^{2}$~\cite{deAraujo95p627};
rhombohedral BiFeO$_3$ could possess a high polarization of $\approx$100$~\mu$C/cm$^2$ and a coercive field in the range of 100--1500~kV/cm~\cite{Yun04p647,Wang03p1719}. The origin of polarization reversal in ZMO remains an open question, except the hypothesis that the Mg dopants may flatten the overall ionic potential energy landscape, similar to that in Sc-doped AlN. It is worthy to establish an atomistic-level switching mechanism that may help the development and optimization of ferroelectric ZMO and similar systems. To address this question, we carry out first-principles density functional theory (DFT) calculations to investigate the switchability of ZMO over a broad composition range. 

\section{Computational Methods}
First-principles DFT calculations are performed using \texttt{QUANTUM ESPRESSO} ~\cite{Giannozzi09p395502, Giannozzi17p465901} package (v6.4.1) with Garrity-Bennett-Rabe-Vanderbilt (GBRV) ultrasoft pseudopotentials~\cite{Garrity14p446}. The exchange-correlation functional is treated within the generalized gradient approximation of Perdew-Burke-Ernzerhof revised for solids (PBEsol) type as PBEsol improves equilibrium properties of densely packed solids~\cite{Perdew96p3865}.
Particularly, PBEsol gives more accurate predictions than PBE on lattice constants of ferroelectrics such as PbTiO$_3$~\cite{Zhao08p184109}. We compared the lattice constants of MgO and ZnO predicted by LDA, PBE, and PBEsol and found that the PBEsol values agree well with experimental results.
Structural parameters of unit cells of ZnO, Zn$_{0.5}$Mg$_{0.5}$O, MgO, and PbTiO$_3$ are optimized using an $8 \times 8 \times 8$ Monkhorst-Pack $k$-point grid centered on the $\Gamma$ point for Brillouin zone sampling, an energy convergence threshold of 10$^{-7}$ Ry, a force convergence threshold of 10$^{-6}$ Ry/Bohr, and a stress threshold of 0.5 kBar. The plane-wave cutoff is set to 80 Ry and the charge density cutoff is set to 600 Ry, respectively. The phonon spectrum is calculated in the framework of density functional perturbation theory (DFPT)~\cite{Baroni87p1861, Gonze95p1086} using a $2 \times 2 \times 2$ $q$-point grid centered on the $\Gamma$ point (\ie, the $q$-point grid includes the $\Gamma$ point). 
The non-analytical correction (NAC) is added to the $\Gamma$ point dynamical matrix file using the calculated dielectric constant and Born effective charges. The real-space interatomic force constants (IFCs) with the NAC correction are then obtained using the code \texttt{q2r.x}. The phonon frequencies at a generic $q$ point are computed by Fourier interpolating the real-space IFCs. The type of acoustic sum rule is set to \texttt{crystal}. 

For a given composition, the energy of a Zn$_{1-x}$Mg$_x$O supercell depends on the arrangement of Zn and Mg ions. As the configuration space increases rapidly with the size of the supercell, it is computationally demanding to exhaust all possible configurations with DFT calculations. We performed test calculations with a $2 \times 2 \times 2$ ZMO supercell of 32 atoms containing two Mg atoms. By comparing the energies of configurations with different Mg-Mg separations, we found that Mg dopants prefer to stay apart. Therefore, at each $x$, we 
construct $3 \times 3 \times 2$ supercells with configurations generated using the \texttt{Supercell} program (v1.0.0) ~\cite{Okhotnikov16p17}, and select five configurations of Zn$_{1-x}$Mg$_x$O with Mg dopants mostly homogeneously distributed in the supercell. These configurations are fully optimized with the plane-wave cutoff, charge density cutoff, and $k$-point grid set to 60 Ry, 500 Ry, and $4 \times 4 \times 4$, respectively. The energy variation among different configurations of the same $x$ is found to be small (within 2~meV/supercell), and the lattice constants of different configurations are very close with a standard deviation smaller than 0.003~\AA.
The lowest-energy configuration is then used for polarization calculations. Nevertheless, we are not claiming the lowest-energy configuration is the experimental configuration given the limited configuration space sampled here. The minimum energy paths (MEPs) of polarization reversal under the clamped-strain condition are determined using the nudged elastic band (NEB) method while switching paths allowing strain relaxations are identified with the variable-cell NEB (VCNEB) technique, both implemented in the \texttt{USPEX} code~\cite{Oganov06p244704,Lyakhov13p1172,Oganov11p227}(v10.3). Each path is constructed by at least 10 images. When the root-mean-square forces are lower than 0.03 eV/\AA~on images or the energy barrier remains unchanged for successive 10 steps, VCNEB calculations are considered converged. All structural files and some representative input files for DFT calculations are uploaded to a public repository~\cite{L39_data}.

The chemical bonding situations in ZnO and MgO are analyzed with the projected Crystal Orbital Hamilton Population (pCOHP) method implemented in \texttt{LOBSTER} (v3.2.0)~\cite{Dronskowski93p8617, Maintz16p1030}. The pCOHP is an energy-resolved partitioning technique that transforms the delocalized electronic structure computed in the reciprocal space into the real space in an energy-resolved form. After a self-consistent electronic structure calculation using \texttt{QUANTUM ESPRESSO} with the same settings detailed above and projector-augmented wave pseudopotentials taken from PseudoDOJO~\cite{Jollet14p1246}, the pCHOP method is used to decompose the band-structure energy into specified interatomic interactions (Zn-O in ZnO and Mg-O in MgO along the $c$ axis).

\section{Results and Discussions}
\subsection{Structure}
\begin{figure}  
\centering
\includegraphics[width=1\textwidth]{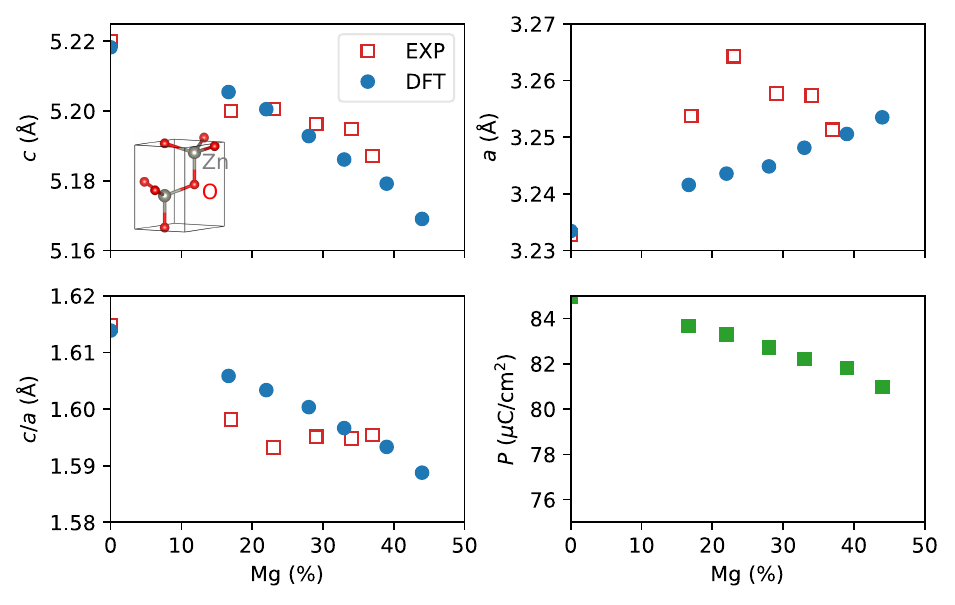} 
\caption{Compositional dependence of lattice parameters ($a$, $c$, and $c/a$) and bulk polarization of Mg-doped ZnO predicted by PBEsol. The inset shows the unit cell of wurtzite ZnO. The red squares are experimental data taken from ref.~\cite{Ferri21p044101}.}
\label{LC}
\end{figure}

For the two-end members of ZMO systems, the PBEsol lattice parameters are $a_0^{\rm ZnO}=3.233$~\AA, $c_0^{\rm ZnO}=5.218$~\AA~for ZnO, and $a_0^{\rm MgO}=3.302$~\AA, $c_0^{\rm MgO}=5.021$~\AA~for MgO, both agreeing well with experimental values, $a_{\rm exp}^{\rm ZnO}=3.25$~\AA, $c_{\rm exp}^{\rm ZnO}=5.21$~\AA~for ZnO~\cite{Sawai10p063541}, and $a_{\rm exp}^{\rm MgO}=3.28$~\AA, $c_{\rm exp}^{\rm MgO}=5.10$~\AA~for MgO~\cite{Jang12p073503}. 
Figure~\ref{LC} reports the DFT-optimized structural parameters and polarization magnitudes of ZMO as a function of Mg percentage ($x$). With increasing $x$, ZMO has $c$ reducing and $a$ increasing; the lattice parameter ratio $c/a$ becomes smaller with increasing MgO alloying. The compositional dependencies of $a$, $c$, and $c/a$ obtained from DFT calculations are consistent with the latest experimental data (red squares in Fig.~\ref{LC})~\cite{Ferri21p044101}.
In particular, the differences between theoretical and experimental values of $c/a$ are within 1\% over the whole Mg concentration range.
Considering the remarkable accuracy of PBEsol on predicting structural properties of ZnO and MgO, we believe the PBEsol values of $a$, $c$, and $c/a$ can be used as reference intrinsic values for defect-free single crystals of ZMO. Because the experimental values of $a$ are $\approx$0.5\% larger than theoretical values for ZMO systems with 16\%--33\% Mg (Fig.~1), we deduce that the synthesized polycrystalline thin films of ZMO likely experienced residual in-plane tensile strains.
Additionally, DFT predicts a remnant polarization of 85~$\mu$C/cm$^2$ for pure ZnO, and a high Mg alloying of 44\% only slightly reduces the magnitude to 81~$\mu$C/cm$^2$. 
Interestingly, Ferri~\etal reported even larger polarization values exceeding $100$~$\mu$C/cm$^2$, nearly 20\% higher than DFT values, in synthesized Zn$_{1-x}$Mg$_{x}$O thin films for $x$ between 0.30 and 0.37. Given that an ideal single crystal without defects is assumed in DFT calculations, the theoretical polarization value thus corresponds to the intrinsic contribution and often serves as an upper bound for experimentally measurable polarization. At this stage, we suggest further studies are needed to resolve the origin of giant remnant polarization in Mg-substituted ZnO thin films reported in ref.~\cite{Ferri21p044101}. 

\begin{figure}  
\centering
\includegraphics[width=0.8\textwidth]{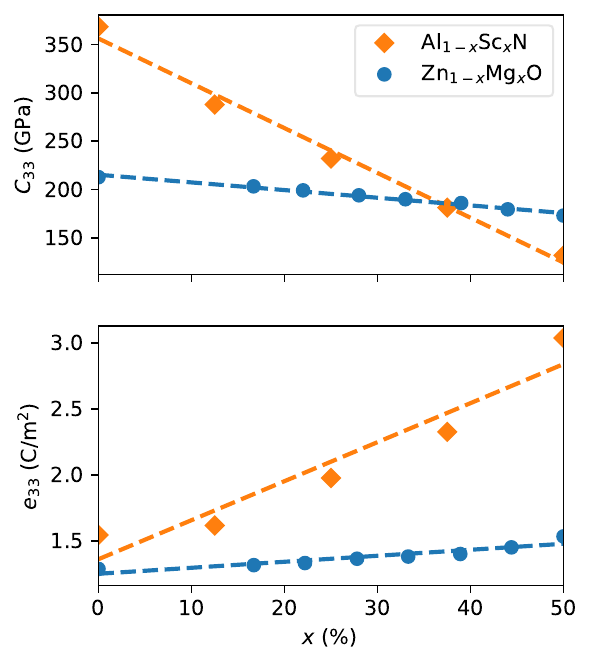} 
\caption{Compositional dependence of elastic constants $C_{33}$ and piezoelectric constants $e_{33}$ of Al$_{1-x}$Sc$_x$N and Zn$_{1-x}$Mg$_x$O solid solutions. The data for Al$_{1-x}$Sc$_x$N is taken from ref.~\cite{Tasnadi10p137601}.}
\label{elastic}
\end{figure}

The magnitude of the polarization often positively correlates with the coercive field of ferroelectrics~\cite{Shin07p881, Liu16p360}. The finding that ZMO with a high content of MgO has nearly the same polarization as pure ZnO seems to suggest ZMO is almost as ``hard" as ZnO. This is supported by the computed elastic stiffness
constants $C_{33}$ for ZMO alloys with different doping concentrations. As shown in Fig.~\ref{elastic}, in the case of another doping-induced ferroelectric system, Al$_{1-x}$Sc$_x$N, the significant elastic softening is evident as $C_{33}$ reduces drastically from 368.2~GPa in AlN to 131.8~GPa in Al$_{0.5}$Sc$_{0.5}$N. In comparison, the value of $C_{33}$ only changes from 212.7 GPa in ZnO to 172.9 GPa in Zn$_{0.5}$Mg$_{0.5}$O~\cite{Tasnadi10p137601}. Similarly, unlike Al$_{1-x}$Sc$_x$N which has the piezoelectric constant $e_{33}$ increases rapidly with the amount of Sc, the piezoelectric response in ZMO systems shows a weak dependence on Mg concentration. 

We examine further the cohesive energies of ZMO systems based on a simple ionic model that treats atoms as fixed point charges. The Madelung energies ($E_{\rm M}$) of polar ZnO, Zn$_{0.5}$Mg$_{0.5}$O, and MgO are computed using Bader charges obtained by partitioning the DFT charge density grid into Bader volumes~\cite{Tang09p084204}. As reported in Table~\ref{Em}, the Bader charge of Mg in MgO is larger than that of Zn in ZnO, and the Madelung energy of wurtzite MgO turns out to be much more negative than that of ZnO. Within the ionic model, MgO appears to have much stronger ionic bonds than ZnO. These results already indicate that the Mg alloying does not induce ``ionic potential softening".

\begin{table}[ht]
\centering
\caption{Bader charges of ions in wurtzite ZnO, Zn$_{0.5}$Mg$_{0.5}$O, and MgO, and their corresponding Madelung energies ($E_{\rm M}$) in eV/atom. There are two unique oxygen atoms in the 4-atom unit cell of Zn$_{0.5}$Mg$_{0.5}$O.}
\label{Em}
\begin{tabular}{cccc}
\hline
\hline
 &{ZnO} & {Zn$_{0.5}$Mg$_{0.5}$O}
 &{MgO}\\
\hline
Zn& $+1.19$ &$+1.14$ &-\\ 
Mg& $-$ &$+1.69$ &$+1.68$\\ 
O& $-1.19$ &$-1.32$ ($-1.52$) &$-1.68$\\ 
\hline
$E_{\rm M}$ & -8.48&-12.23 &-16.86 \\
\hline
\hline
\end{tabular}
\end{table}

\subsection{Switchability gauged by polar-nonpolar energy difference}
A spontaneous polarization switchable by an external electric field is a hallmark feature of ferroelectricity. The energy difference between the ferroelectric (FE) phase and the paraelectric (PE) reference phase, $\Delta U = U_{\rm PE} - U_{\rm FE} $, was often used to gauge the switchability as this quantity is easily accessible via DFT calculations~\cite{Bennett08p17409}. Though the structure of the polar phase for a ferroelectric is often the ground state and unambiguous, the choice of the corresponding nonpolar reference structure can be subtle. For example, the zincblende structure (space group $F\bar{4}3m$) has been used as a nonpolar reference to determine the spontaneous polarization constants for the wurtzite structure~\cite{FabiopR10024,Jang12p73503}. However, Dreyer~{\etal}~pointed out that the layered hexagonal ($h$) structure in the space group of $P6_3/mmc$ should be adopted in the Berry phase approach for polarization calculation~\cite{Dreyer16p21038}. Moreover, when assuming a homogeneous switching process, the $h$-phase is a natural intermediate nonpolar state bridging two oppositely polarized states. For these two reasons, we choose the $h$-phase as the nonpolar reference in this work.

\begin{figure}  
\centering
\includegraphics[width=1\textwidth]{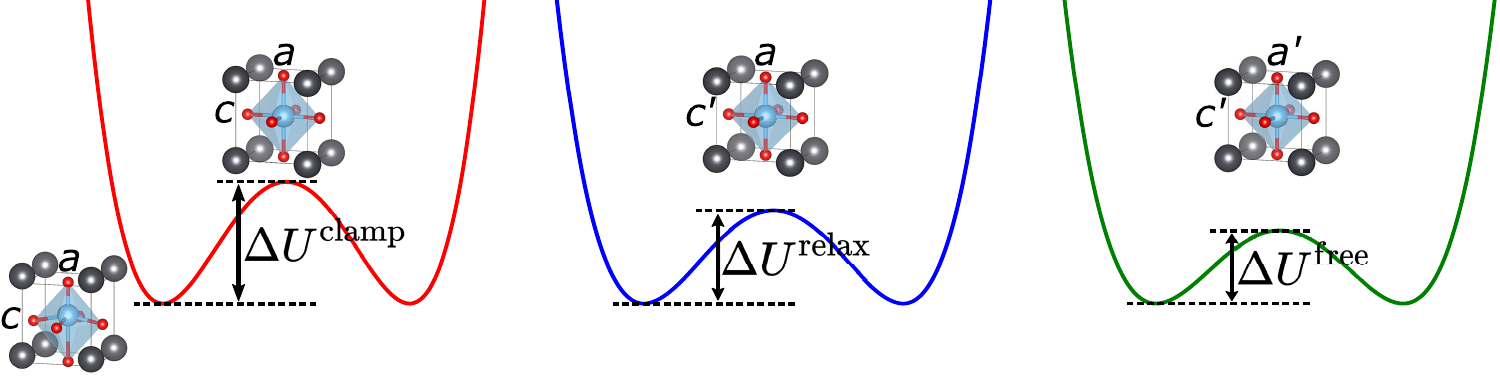} 
\caption{Schematics of different definitions of polar-nonpolar energy differences using PbTiO$_3$ as an example.}
\label{PTO}
\end{figure}

Another subtlety is due to the sensitive dependence of the soft-mode potential energy surface on the strain and volume~\cite{Cohen92p136}. Taking the prototypical ferroelectric, tetragonal PbTiO$_3$ (space group $P4mm$), as an example, one may construct a zero-polarization phase by displacing atoms of the ferroelectric phase along the soft mode. The lattice constants of the nonpolar phase nevertheless can have different choices, leading to different definitions of $\Delta U$ (Fig.~\ref{PTO}). One can clamp the lattice constants to the values of the ground-state ferroelectric phase while fixing atoms at the high-symmetry sites, thus creating a nonpolar phase in the space group of $P4/mmm$; the corresponding ferroelectric-paraelectric energy difference is denoted as $\Delta U^{\rm clamp}$ (Fig.~\ref{PTO}, left). The second choice is to relax the lattice constant along the polarization direction ($c$-axis for PbTiO$_3$) at zero in-plane strain (by fixing $a$-axis) while all the internal atomic coordinates are fixed at the high-symmetry sites; the energy barrier between this nonpolar variant and the ferroelectric phase is denoted as $\Delta U^{\rm relax}$ (Fig.~\ref{PTO}, middle) and is likely relevant for polarization switching in thin films subjected to in-plane clamping from substrates. Finally, one may obtain the lowest-energy cubic phase in the space group of $Pm\bar{3}m$ with the resulting energy difference denoted as $\Delta U^{\rm free}$ (Fig.~\ref{PTO}, right). Regardless of the definitions, the magnitudes of $\Delta U^{\rm clamp}$, $\Delta U^{\rm relax}$, and $\Delta U^{\rm free}$ reflect the polarization switchability under different mechanical boundary conditions, and their varying magnitudes essentially manifest the coupling strength between ferroelectric distortions and strain/volume. It is expected that $\Delta U^{\rm clamp}>\Delta U^{\rm relax}>\Delta U^{\rm free}$ as more degrees of freedom are allowed to relax in sequence; indeed, for PbTiO$_3$, we find $\Delta U^{\rm clamp}=67.8$~meV/atom, $\Delta U^{\rm relax}=23.6$~meV/atom, and $\Delta U^{\rm free}=17.7$~meV/atom using PBEsol. We will use these three types of $\Delta U$ to evaluate the polarization switchability of ZMO systems. At each $x$, we first obtain the optimized lattice constants of the polar wurtzite phase and then compute the energies of three nonpolar variants and the corresponding $\Delta U$ as discussed above. 

\begin{figure}  
\centering
\includegraphics[width=1\textwidth]{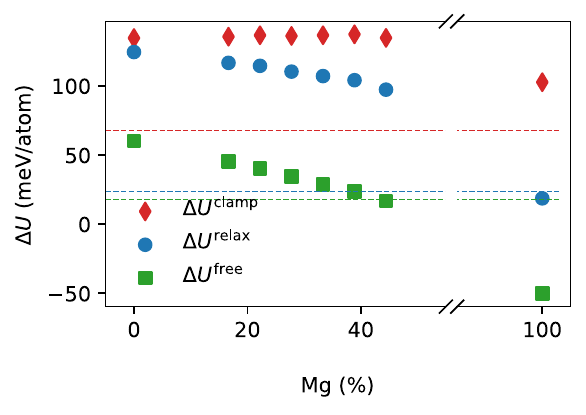} 
\caption{Polar-nonpolar energy difference ($\Delta U$) computed using different nonpolar variants (see discussions in the main text) as a function of Mg content. The dashed red, blue, and green lines mark the values of $\Delta U^{\rm clamp}$, $\Delta U^{\rm relax}$, and $\Delta U^{\rm free}$ of PbTiO$_3$, respectively. }
\label{dU}
\end{figure}

\begin{figure}  
\centering
\includegraphics[width=1\textwidth]{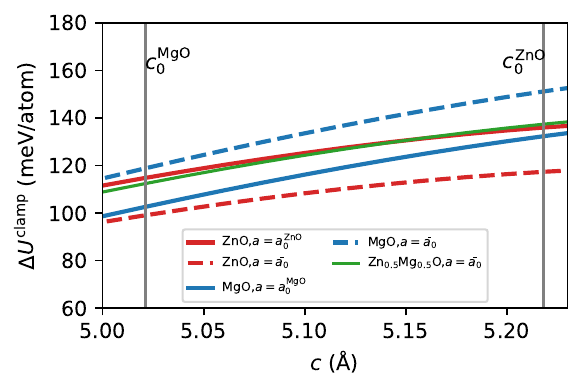} 
\caption{Strain-dependence of $\Delta U^{\rm clamp}$ in ZnO, MgO, and Zn$_{0.5}$Mg$_{0.5}$O with $a$-axis fixed to representative values. All calculations are performed with a 4-atom unit cell. $a_0^{\rm ZnO}=3.233$~\AA, $c_0^{\rm ZnO}=5.218$~\AA~for ZnO, and $a_0^{\rm MgO}=3.302$~\AA, $c_0^{\rm MgO}=5.021$~\AA~for MgO, $\bar{a}_0=3.268$~\AA.
}
\label{compareZnOMgO}
\end{figure}

Figure.~\ref{dU} reports the $x$-dependent $\Delta U^{\rm clamp}$, $\Delta U^{\rm relax}$, and $\Delta U^{\rm free}$ of ZMO systems. We find that $\Delta U^{\rm clamp}$ has a rather weak dependence on $x$. Surprisingly, even at a high Mg concentration of 44.4\%, $\Delta U^{\rm clamp}$ remains large, $\approx135$~meV/atom, nearly the same as that of pure ZnO. The value of $\Delta U^{\rm clamp}$ for MgO is also significant, $\approx103$~meV/atom. 
These results confirm that Mg alloying does not substantially soften the ionic potential energy landscape in the absence of strain relaxation. We thus explore the relationship between $\Delta U^{\rm clamp}$ and strain by performing model calculations for unit cells of ZnO, MgO, and Zn$_{0.5}$Mg$_{0.5}$O. As shown in Fig.~\ref{compareZnOMgO}, when the $a$-axis is fixed to the ground-state value ($a_0^{\rm ZnO}$=3.233~\AA~and $a_0^{\rm MgO}$=3.302~\AA~for MgO), $\Delta U^{\rm clamp}$ decreases with reducing $c$ in both MgO and ZnO, and MgO appears to be slightly ``softer" than ZnO at the same $c$ (see solid lines in Fig.~\ref{compareZnOMgO}). However, when the $a$-axis is fixed to the averaged value of $a_0^{\rm ZnO}$ and $a_0^{\rm MgO}$ (denoted as $\bar{a}$), that is equivalent to applying in-plane compressive strain to MgO and tensile strain to ZnO, MgO turns out to be much ``harder" than ZnO at the same $c$ (see dashed lines in Fig.~\ref{compareZnOMgO}). Because of this cancellation effect, the magnitude of $\Delta U^{\rm clamp}$ of Zn$_{0.5}$Mg$_{0.5}$O is nearly identical to that of pure ZnO over a wide range of $c$ values when $a=\bar{a}_0$. Results from these model calculations indicate that the conventional argument that Mg alloying may ``soften" ZnO is overly simplified.

\begin{figure}  
\centering
\includegraphics[width=0.8\textwidth]{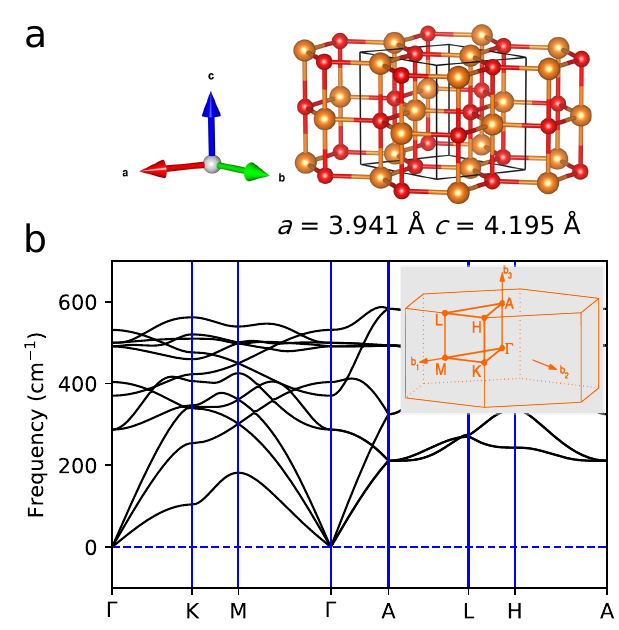} 
\caption{(a) Schematic of hexagonal phase of MgO in the space group of $P6_3/mmc$. Mg atoms are denoted by orange balls and O atoms are denoted by red balls. (b) Phonon spectrum along high-symmetry lines of the Brillouin zone of the hexagonal unit cell. }
\label{HMgO}
\end{figure}

In comparison, both the magnitudes of $\Delta U^{\rm relax}$ and $\Delta U^{\rm free}$ decrease with increasing Mg concentrations, and $\Delta U^{\rm free}$ is much smaller than $\Delta U^{\rm relax}$ and $\Delta U^{\rm clamp}$ at the same $x$ (Fig.~\ref{dU}). For example, at $x=0.444$, $\Delta U^{\rm clamp}$ is $135$~meV/atom, $\Delta U^{\rm relax}$ is $97.3$~meV/atom, while $\Delta U^{\rm free}$ is only 16.7~meV/atom, even lower than that of PbTiO$_3$ ( 17.7~meV/atom). We attribute the small magnitude of $\Delta U^{\rm free}$ of ZMO to the hexagonal phase of MgO ($h$-MgO, see Fig.~\ref{HMgO}a) as this nonpolar phase is much lower in energy than the polar $wz$-MgO, $\Delta U^{\rm free}=-50.2$~meV/atom. In addition, the phonon spectrum of $h$-MgO has no imaginary frequencies (Fig.~\ref{HMgO}b), indicating this phase is dynamically stable within a harmonic approximation at 0~K.
We propose that the low-energy $h$-MgO plays a key role in the emergence of switchable polarization in ZMO systems. 

\subsection{Energy-polarization landscape obtained with constrained-$P$ optimizations}
\begin{figure}  
\centering
\includegraphics[width=0.8\textwidth]{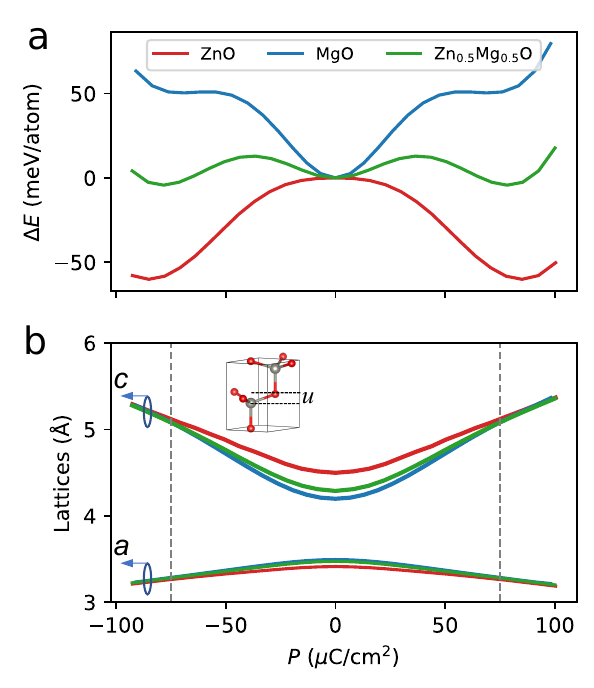} 
\caption{(a) Energy landscape as a function of polarization ($P$) obtained with constrained-$P$ optimizations. The nonpolar state is chosen as the zero energy point for each system. (b) Lattice constants as a function of $P$.}
\label{PES}
\end{figure}

The finding that $\Delta U^{\rm free}\ll \Delta U^{\rm clamp}$ in Zn$_{1-x}$Mg$_{x}$O solid solutions indicates that the strain relaxation is crucial for the observed ferroelectricity in ZMO systems. It will be helpful to construct the polarization-energy landscape $E(P)$ that takes the strain relaxation effect into account. 
This demands the optimization of crystal lattice parameters and atomic positions at a fixed $P$, a nontrivial task that is not easily implementable within DFT~\cite{Stengel09p304}. In recognition that the polarization of the wurtzite phase strongly correlates with the separation (denoted as $u$ in the inset of Fig.~\ref{PES}b) between the Zn (Mg) layer and O layer along the $c$ axis, we can optimize the lattice constants and other internal degrees of freedom at a fixed value of $u$. The polarization of such optimized structure is then computed with the Berry phase approach. This protocol allows us to construct $E(P)$ profiles for ZnO, Zn${_{0.5}}$Mg${_{0.5}}$O, and MgO, respectively. As shown in Fig.~\ref{PES}a, ZnO has a larger barrier separating two oppositely polarized states, while MgO strongly favors the nonpolar state. Notably, the $E(P)$ profile of Zn$_{0.5}$Mg$_{0.5}$O is a shallow triple well and is very close to the equal-weighted sum of the energy profiles of ZnO and MgO. The strain relaxation effect drastically reduces the barrier in Zn$_{0.5}$Mg$_{0.5}$O. We plot the changes in lattice constants along $E(P)$ profiles. It is interesting that MgO and ZnO have comparable lattice constants at the same $P$ in the strongly polarized region ($|P|>75~\mu$C/cm$^2$, highlighted as dashed gray lines in Fig.~\ref{PES}b). We suggest that such ``lattice-polarization matching" between polar ZnO and MgO is beneficial for the formation of homogeneous solid solutions of ZMO: adjacent ZnO and MgO unit cells adopting the same polarization will possess similar lattice constants thus minimizing the elastic energy penalty.\\

\subsection{Switchability gauged by NEB and VCNEB}

\begin{figure}  
\centering
\includegraphics[width=0.6\textwidth]{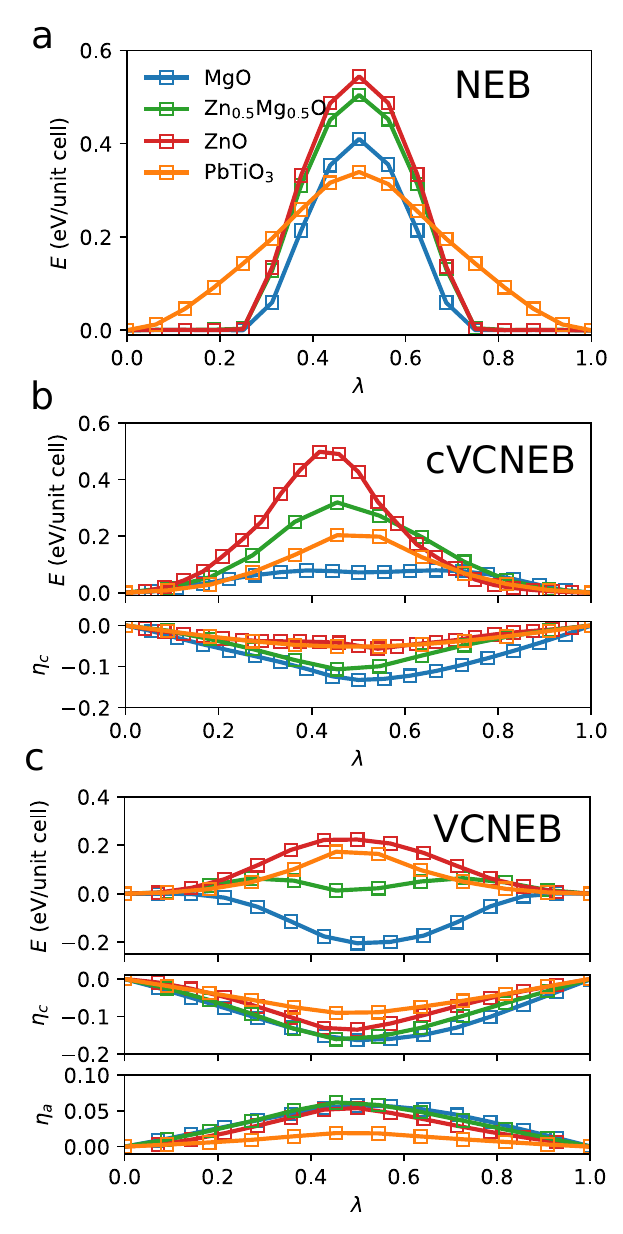} 
\caption{Minimum energy paths (MEPs) of polarization reversal identified with (a) NEB, (b) cVCNEB with fixed in-plain strain, and (c) VCNEB. $\eta_a$ and $\eta_c$, defined as $a/a_0-1$ and $c/c_0-1$, respectively, are the strain variations along the $a$-axis and $c$-axis along the MEPs; $a_0$ and $c_0$ are the lattice constants of ground-state polar phases.}
\label{NEBall}
\end{figure}

The polarization reversal paths identified using NEB and VCNEB methods further emphasize the importance of the strain relaxation effect. 
We map out the MEP of polarization reversal using the NEB method in which the lattice constants of all the images along the switching pathway are fixed to the values of the ground-state ferroelectric phase. Figure.~\ref{NEBall}a presents the MEPs for unit cells of ZnO, Zn$_{0.5}$Mg$_{0.5}$O, MgO, and PbTiO$_3$, respectively. The barriers obtained with NEB, $\Delta U^{\rm NEB}$, are 129 meV/atom for ZnO, 124 meV/atom for Zn$_{0.5}$Mg$_{0.5}$O, 102 meV/atom for MgO, and 65.2 meV/atom for PbTiO$_3$, all comparable to their respective $\Delta U^{\rm clamp}$ in magnitude. The finding that ZnO, Zn$_{0.5}$Mg$_{0.5}$O, and MgO have 
switching barriers comparably higher than PbTiO$_3$ corroborates our hypothesis that even a high Mg concentration is not enough to soften the energy landscape of polarization reversal in the absence of strain relaxation. 

We then introduce the relaxation effect into NEB calculations. By analogy with $\Delta U^{\rm relax}$, another sets of MEPs are obtained by fixing in-plane lattice constant $a$ while allowing lattice constant $c$ to change along the switching pathway. Such constrained-variable-cell NEB, denoted as cVCNBE, allows the modeling of polarization reversal processes in thin films where the in-plane clamping effect is important. As shown in Fig.~\ref{NEBall}b, the freedom to vary strain along $c$ during the switching has little impact on the polarization reversibility of ZnO: the cVCNEB switching barrier, $\Delta U^{\rm cVCNEB}$, is 125 meV/atom for ZnO, nearly identical to $\Delta U^{\rm NEB}$. In comparison, MgO has a marked different MEP when the strain is allowed to relax: the polarization reversal only needs to overcome a small enthalpy barrier of $\approx$ 20~meV/atom, even lower than that in PbTiO$_3$. The barrier height in Zn$_{0.5}$Mg$_{0.5}$O reduces to $\approx$80~meV/atom. We compare the strain change $\eta_c$, defined as $c/c_0-1$, along the MEPs of cVCNEB. The switching processes in MgO ad Zn$_{0.5}$Mg$_{0.5}$O involve much larger changes in $\eta_c$ than those in PbTiO$_3$ and ZnO. The effect of temperature on the ferroelectricity of ZMO thin films remains unexplored in experiments as this ferroelectric system is only discovered recently. 
Since the DFT-computed switching barrier at zero Kelvin often positively correlates with the magnitude of Curie temperature ($T_C$), it is expected that $T_C$ of ferroelectric ZMO thin films will be higher than that of PbTiO$_3$-based thin films since the former has a larger switching barrier.

Finally, we compute the MEPs with VCNEB that allows both $c$ and $a$ to vary during the switching, and the results are reported in Fig.~\ref{NEBall}c. These MEPs largely resemble the $E(P)$ profiles obtained with constrained-$P$ optimizations shown in Fig.~\ref{PES}. For example, the MEP of MgO confirms that the nonpolar $h$-MgO has lower energy than the polar $wz$-MgO. Importantly, the switching in Zn$_{0.5}$Mg$_{0.5}$O with a barrier of only 16~meV/atom now becomes highly feasible when all strains are allowed to relax. The finding that NEB and VCNEB predict very different polarization reversal barriers for Zn$_{0.5}$Mg$_{0.5}$O has important implications.
The main difference between NEB and VCNEB is that VCNEB allows lattice constants to change during solid-solid transformations. The NEB method is an efficient and robust approach for identifying the minimum energy path connecting two given structures, and it has been successfully employed to study many problems such as molecular chemical reactions~\cite{Ke21p10860}, surface adsorptions~\cite{Ciobica03p3808}, and defect migration~\cite{Erhart06p115207}. However, the application of the NEB method demands the initial and final structures possessing the same lattice constants. For this reason, the NEB method cannot quantify the impact of possible strain relaxation during the ferroelectric switching. In comparison, the VCNEB method fully captures the possible strain relaxation effect. 
The much lower switching barrier obtained with VCNEB than that obtained with NEB is a clear indicator that the strain variation is a key enabler of the switchable polarization in ZMO systems. 

\subsection{Comparison between $h$-ZnO and $h$-MgO}
\begin{figure}  
\centering
\includegraphics[width=0.8\textwidth]{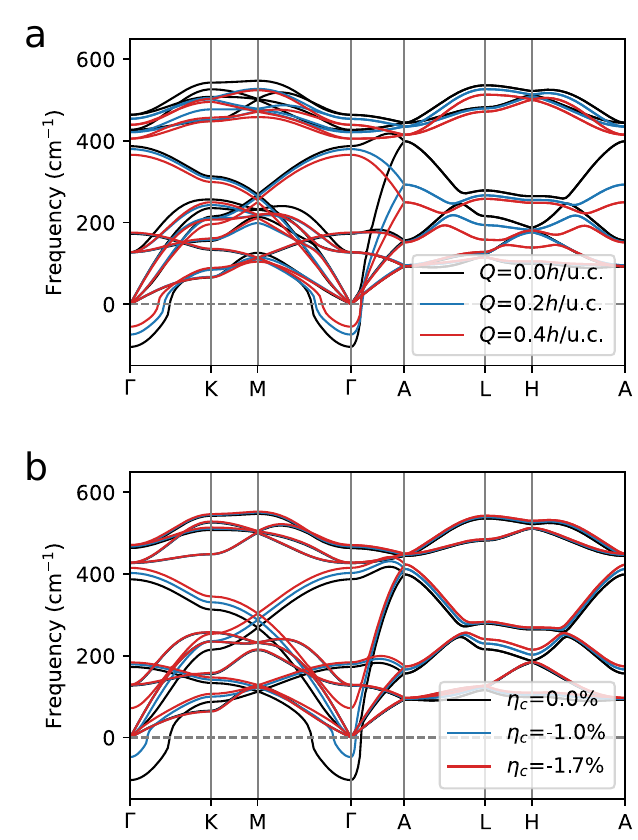} 
\caption{Phonon dispersion relationships of (a) hole-doped $h$-ZnO, (b) strained $h$-ZnO. The NAC is not applied in these calculations.}
\label{ZnOPhonon}
\end{figure}

Despite the nearly identical ionic radii of Zn$^{2+}$ and Mg$^{2+}$, 0.57~\AA~versus 0.6~\AA, $h$-ZnO and $h$-MgO are drastically different. As shown in Fig.~\ref{HMgO}, $h$-MgO is dynamically stable as the phonon dispersion curve has no modes with imaginary frequencies. In comparison, $h$-ZnO is dynamically unstable, exhibiting imaginary phonon modes in the region close to the center of the Brillouin zone ($\Gamma$, see Fig.~\ref{ZnOPhonon}). The most unstable mode is at $\Gamma$, which is a transverse optic (TO) mode, dominated by the Zn displacement against oxygen atoms of the same layer along the $c$ axis, that freezes in to give the polar wurtzite phase. It is thus natural to ask about the differences between the hexagonal phases of ZnO and MgO. Specifically, why the nonpolar $h$-ZnO is less stable while $h$-MgO being more stable than their polar counterparts? 
We address this question from two different perspectives. 

First, the origin of ferroelectricity in perovskite oxides is often understood as a delicate balance between the long-range electrostatic dipole-dipole interaction and the short-range repulsion, with the former favors the polar phase while the latter favors the nonpolar phase~\cite{Cohen92p136}. We find that in the case of $h$-ZnO, the long-range Coulomb interaction dominates over the short-range repulsion through a set of model calculations. As shown in Fig.~\ref{ZnOPhonon}a, with increased hole doping that screens the long-range electrostatic interaction, a clear mode hardening appears at $\Gamma$. Furthermore, the short-range repulsion can be enhanced by reducing the $c$ axis. As expected, a compressive strain along the $c$ axis ($\eta_c$) drives the soft mode harder, and $h$-ZnO becomes dynamically stable at $\eta_c=-1.7$\% (Fig.~\ref{ZnOPhonon}b). This is also in line with the finding that $h$-MgO has a smaller $c$ than $h$-ZnO. 

\begin{figure}  
\centering
\includegraphics[width=1\textwidth]{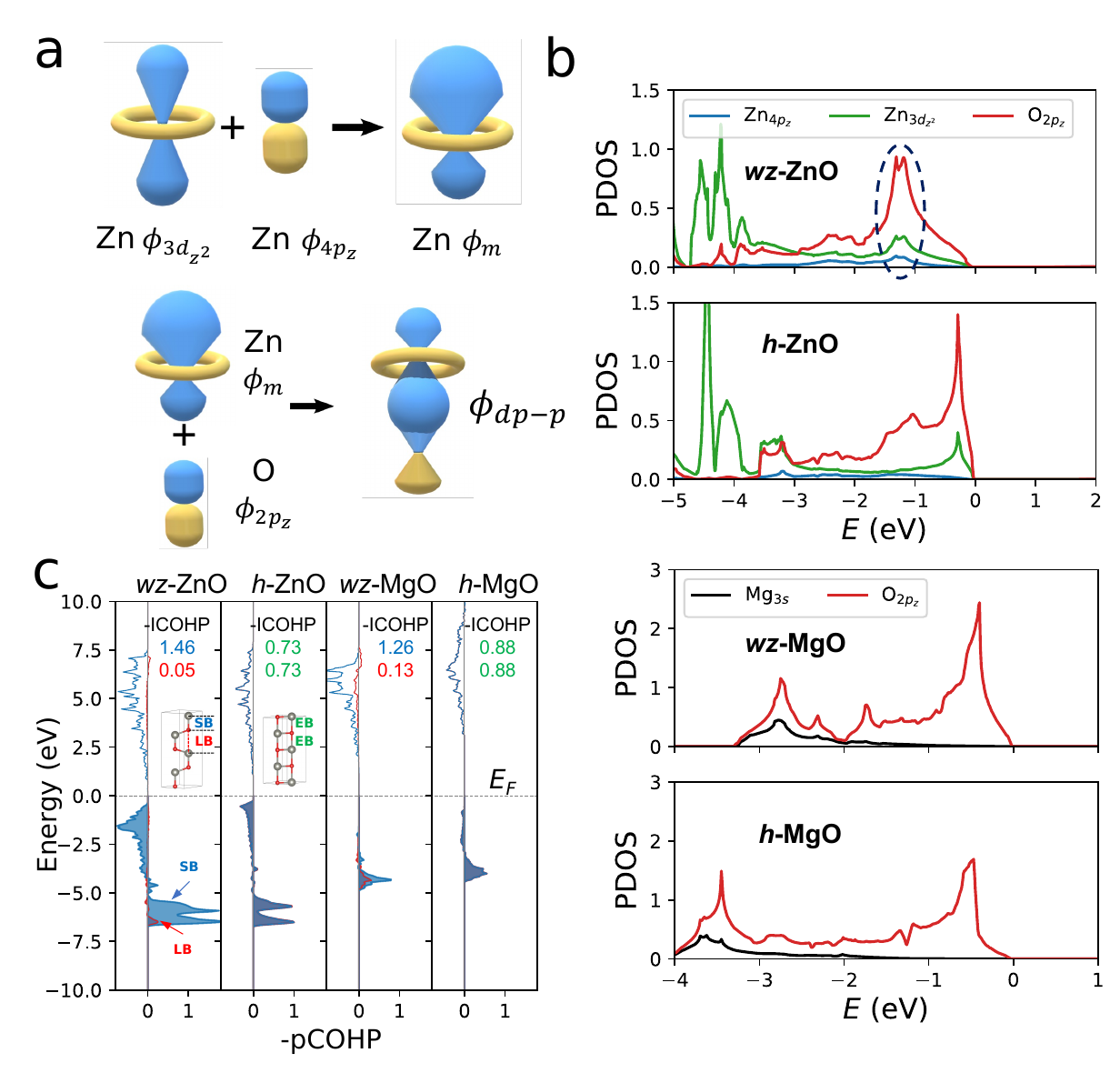} 
\caption{(a) Schematics of a sequential orbital interaction mechanism that explains the origin of inversion-symmetry breaking in $wz$-ZnO. The self-mixing of Zn-$3d_{z^2}$ and Zn-$4p_z$ orbitals results in an orbital $\phi_m$. The asymmetric $\phi_m$ can hybridize with the O-$2p_z$ orbital. (b) Projected density of states in $wz$-ZnO, $h$-ZnO, $wz$-MgO, and $h$-MgO from PBE calculations. The dashed oval highlights the $dp$-$p$ hybridization. (c) Calculated pCOHP curves for Zn(Mg)-O pairs along the $c$ axis with values of $-$ICOHP of SB, LB, and EB in blue, red, and green, respectively. The insets illustrate the SB and LB in $wz$-ZnO and two EBs in $h$-ZnO.}
\label{COHP}
\end{figure}

The thermodynamic stability of $wz$-ZnO over $h$-ZnO can also be understood by analyzing the orbital interactions and Zn-O bonding strengths. It is well established that ferroelectric perovskites often involve $d^0$ cations such as Ti$^{4+}$ so that the empty $d$-orbitals tend to hybridize with O-$2p$ orbitals to induce spontaneous off-center displacements. Oak~\etal proposed a covalent bonding mechanism to explain the hexagonal ferroelectricity in InMnO$_3$ where the In$^{3+}$ ion has a fully occupied 4$d$ orbital: it is the intra-atomic 4$d_{z^2}$-5$p_z$ orbital mixing of In followed by asymmetric $4d_{z^2}$(In)-$5p_z$(O) bonding along the $c$ axis that drives the break of inversion symmetry~\cite{Oak11p047601}. Similarly, Lee~\etal pointed out that the Zn-$3d$ orbitals are not capable of forming hybridized orbitals with the O-$2p_{z}$ orbitals because the Zn-$3d$ orbitals are fully filled~\cite{Lee15p7857}. To resolve this puzzle, a sequential orbital interaction mechanism has been proposed~\cite{Lee15p7857} and is depicted in Fig.~\ref{COHP}a. The self-mixing of Zn-$3d_{z^2}$ and Zn-$4p_z$ orbitals results in an orbital $\phi_m$ characterized with an asymmetric shape along the $c$ axis. This orbital $\phi_m$ is allowed by symmetry to hybridize with the O-$2p_z$ orbital. As shown in Fig.~\ref{COHP}b, the projected density of states (PDOS) reveals overlapping peaks of states of Zn-$3d_{z^2}$, Zn-$4p_z$, and O-$2p_z$ characters at $\approx-1$~eV below the valence band maximum in $wz$-ZnO. In the case of $h$-ZnO, the PDOS of Zn-$3d_{z^2}$ character no longer overlaps with the PDOS of Zn-4$p_z$ character, indicating the lack of Zn $3d_{z^2}$-$4p_z$ orbital self-mixing. The hybridization of $\phi_m$ and O-$2p_z$, denoted as $dp$-$p$ hybridization, is the driving force for the inversion-symmetry breaking that transforms two equal Zn-O bonds (EBs) to one short bond (SB) and one long bond (LB) along the $c$ axis. In $wz$-ZnO, the energy gain of forming a stronger SB outcompetes the energy penalty of forming a weaker LB. We compute projected crystal orbital Hamilton population (pCOHP) curves for Zn-O bonds along the $c$ axis in $wz$-ZnO and $h$-ZnO, respectively. The integrated COHPs (ICOHPs) for the SB is $-1.46$ eV per bond and $-0.05$ eV per bond for the LB in $wz$-ZnO, the total of which is larger (more negative) than the ICOHP sum of two EBs in $h$-ZnO, supporting the argument that Zn-O bonds in $wz$-ZnO are stronger than those in $h$-ZnO. 

In contrary, the electronic configuration of Mg$^{2+}$ is much simpler than Zn$^{2+}$. As shown in the PDOS spectra of $wz$-MgO and $h$-MgO (Fig.~\ref{COHP}b), the states near the Fermi level are dominated by O-$2p$ orbitals. The lack of $dp$-$p$ hybridization turns off the tendency of the off-centering displacement of the Mg$^{2+}$ ion. Indeed, the ICOHP sum of two EBs in $h$-MgO now becomes larger than the ICOHP sum of SB and LB in $wz$-MgO, consistent with the observation that $h$-MgO has lower energy.\\

\section{Conclusion}
The integration of ferroelectric functionalities into integrated circuits demands ferroelectrics that are compatible with the current semiconductor manufacturing process. Recent experimental efforts to make the polarization of silicon-compatible piezoelectrics switchable is a promising strategy to expand the list of CMOS-compatible ferroelectrics. We gauge the polarization switchability in magnesium doped zinc oxide with several key quantities such as polar-nonpolar energy differences and switching barriers computed from NEB and VCNEB based on first-principles density functional theory. The general explanation that Mg dopants flatten the ionic potential energy landscape of Mg-doped ZnO is overly simplified because both MgO and ZnO in the wurtzite phase appear to be comparably hard to switch in the absence of strain relaxation. In particular, unlike Sc-doped AlN, Mg alloying does not induce elastic softening in ZMO systems. We suggest that the hexagonal phase of MgO, being dynamically stable and lower in energy than its wurtzite counterpart, is responsible for the emergence of ferroelectricity in Mg-doped ZnO thin films. The polarization reversal process in ZMO systems is found to involve a large change in strain along the polar direction. This hints that the 180$^{\circ}$ switching of wurtzite ferroelectrics may possess ferroelastic characters, an unusual feature not presented in perovskite ferroelectrics.

\begin{acknowledgments}
J.H., Y.H, and S.L. acknowledge the supports from Natural Science Foundation of Zhejiang Province (2022XHSJJ006) and Westlake Education Foundation. The computational resource is provided by Westlake HPC Center.
\end{acknowledgments}


\bibliography{SL}

\clearpage
\newpage

\clearpage
\newpage

\clearpage
\newpage

\clearpage
\newpage

\clearpage
\newpage

\clearpage
\newpage

\clearpage
\newpage

\clearpage
\newpage

\clearpage
\newpage

\clearpage
\newpage

\end{document}